\documentclass[trackchanges]{aastex701}

\begin{document}

\title{Testing Weak Equivalence Principle with IceCube Event and Blazar}

\author[orcid=0000-0002-9332-5562,sname='Wang']{Tian-Cong Wang}
\affiliation{School of Physics and Astronomy, Beijing Normal University, Beijing 100875, China}
\affiliation{Institute for Frontiers in Astronomy and Astrophysics, Beijing Normal University, Beijing 102206, China}
\email[show]{wangtc@mail.bnu.edu.cn} 

\author{Aleksandra Pi{\'o}rkowska - Kurpas}
\affiliation{
 August Chełkowski Institute of Physics, Faculty of Science and Technology, University of Silesia, 75 Pułku Piechoty 1, Chorzów, 41-500, Poland
 }
 \email{aleksandra.piorkowska@us.edu.pl}

\author{Marek Biesiada}
\affiliation{
 National Center for Nuclear Research, Pasteura 7, Warsaw, 02-093, Poland
}
\email[show]{Marek.Biesiada@ncbj.gov.pl}

\author[orcid=0000-0003-2516-6288,sname='Gao']{He Gao}
\affiliation{School of Physics and Astronomy, Beijing Normal University, Beijing 100875, China}
\affiliation{Institute for Frontiers in Astronomy and Astrophysics, Beijing Normal University, Beijing 102206, China}
\email[show]{gaohe@bnu.edu.cn}

\correspondingauthor{Tian-Cong Wang, Marek Biesiada, He Gao}

\begin{abstract}

Einstein's Weak Equivalence Principle (WEP), the universality of free fall, is a fundamental component of general relativity and other metric theories of gravity. Its validity can be tested through the post-Newtonian parameter $\gamma$, which quantifies the amount of spacetime curvature due to the presence of unit rest mass. In this paper, we use high‑energy neutrino events detected by IceCube and associated with the gamma‑ray blazars TXS 0506+056 and PKS 0735+178 to test the WEP via the Shapiro delay induced by the gravitational potential of Laniakea. We find that violation of the equivalence principle for neutrinos and photons is limited to an accuracy of $10^{-6}$, $10^{-7}$ and $10^{-8}$, representing improvements of one, two, and three orders of magnitude, respectively, over previous constraints obtained from other high‑energy neutrino–blazar associations and up to six orders of magnitude tighter compared to the constraints obtained with MeV neutrinos from SN1987A.

\end{abstract}



\section{\label{sec:Intro}Introduction}

Einstein's theories of Special and General Relativity stand as cornerstones of modern physics, with wide-ranging implications for astrophysics. Consequently, rigorous testing of the fundamental assumptions underlying relativity remains of profound scientific interest. While terrestrial experiments offer valuable insights, the extreme environments present in astrophysical phenomena provide a more suitable testbed for high-precision verification of fundamental physical laws. According to Einstein's Weak Equivalence Principle (WEP), any freely falling, uncharged test body should follow a trajectory that is independent of its internal structure and composition \citep{Misner1973}. This principle is integral to General Relativity and other metric theories of gravity. The most iconic test of the WEP is the E\"{o}tv\"{o}s-type experiment, where the accelerations of two laboratory-scale objects made of different materials are compared in a known gravitational field (see \citealp{Will_1993}, \citealp{Will_2014} and references therein). For macroscopic masses, the validity of the WEP can be tested within the Newtonian framework. However, the dynamics of test particles, such as photons or neutrinos, in a gravitational field require a more sophisticated description than Newtonian mechanics can offer. To this end, the parametrized post-Newtonian (PPN) formalism has been developed to characterize their motion accurately \citep{Thorne_Will_1971, Will_1993}. Under this approach, each gravitational theory that adheres to the WEP is characterized by a specific set of PPN parameters. Consequently, WEP can also be tested using massless (or nearly massless) particles, with any potential deviations being quantified by the PPN parameters ($\alpha$ and $\gamma$) specific to the theory in question.

Specifically, numerous precise constraints on the PPN parameter $\gamma$ have been established by examining the arrival time differences between various particles. These include constraints derived from photons and neutrinos originating from SN1987A in the Large Magellanic Cloud \citep{KraussTremaine_1988, Longo_1988, Bose_1988} or, more recently, from photons observed in different energy bands emitted from the Crab pulsar \citep{YangZhang_2016}, gamma-ray bursts \citep{Gao_2015, Nusser_2016, SangYu_2016} and blazars \citep{Wei_2016}, as well as photons of different frequencies in fast radio bursts \citep{Wei_2015, Nusser_2016, Xing_2019}.  See also \cite{Wu_2017} and \cite{Yang_2017} where constraints on the $\gamma$ parameter were obtained through polarized radio emissions from gamma-ray bursts and fast radio bursts. In the era of successful observations of gravitational wave signals in the large interferometric detectors, conservative limits on the $\gamma$ parameter can also be obtained with double compact object mergers \citep{Wu_2016, Kahya_2016, Abbott_2017, Wei_2017}. A recent analysis also examines constraints on the PPN parameter $\alpha$ from neutrino and photon observations of SN 1987A, providing a test of the universality of free fall for neutrinos \citep{Bhadra_2024}. Given the strong dependence of the above results on the distance between the particle source and the observer, using more distant astrophysical sources may yield more precise bounds on the PPN parameters under consideration.

Blazars rank among the most powerful electromagnetic emitters in the universe. Their rapid brightness variations \citep{Jorstad_2022, Larionov_2020, Bhatta_2020, Gupta_2018, Wehrle_2016, Jorstad_2005} and polarization characteristics \citep{D’Arcangelo_2007, Jorstad_2006, Jorstad_2005, Jannuzi_1993, Kinman_1967} make them key targets for studying black holes, relativistic jets, and high-energy physical processes (see e.g. \citealp{Raiteri_2023, Jorstad_2022, Larionov_2020, Sarkar_2020, Gupta_2018, Raiteri_2017, Negi_2021, Podjed_2024, Bolis_2024, Shkodkina_2025, Liodakis_2021, Zhang_2014, D’Arcangelo_2007} and references therein). 
The hypothesis that blazars may serve as a potential source of high-energy neutrinos has been extensively explored in the literature \cite{Mannheim_1995, Halzen_1997, Mucke_2003, Padovani_2014, Petropoulou_2015, Tavecchio_2015}. The underlying rationale is simple: neutrinos are efficiently produced through the decay of charged pions, preceded by proton-proton or proton-photon interactions (the so-called photo-pion production). Given that material outflow in blazars forms relativistic jets where particles have a chance to accelerate to extreme energies \citep{Urry_1995, Padovani_2017, Kotera_2011}, it thus provides an ideal environment for such processes. Consequently, we expect efficient high-energy neutrino production within the jet through the prompt decay of charged pions.
The first gamma-ray blazar associated with an IceCube neutrino event was TXS 0506+056 \citep{IceCube_2018a, Padovani_2018, Paiano_2018}. This observation provided the first direct evidence that blazars can produce high-energy neutrinos, opening new avenues to test General Relativity and the equivalence principle for TeV neutrinos. The total travel time for neutrinos from this blazar to reach Earth equals the distance divided by their vacuum speed, plus an additional delay due to the non-zero gravitational potential along the line of sight. 
This method has already been adopted by \cite{Wang_2016} to constrain the PPN parameter $\gamma$ using observationally suggested association (about $5\%$ a posteriori probability for chance coincidence translating into about $2\sigma$ confidence correlation) between the PKS B1424-418 blazar's giant flare and the 2 PeV neutrino reported by IceCube Collaboration (IceCube event 35; see the references in \citealp{Wang_2016}). In this paper, we adopt this strategy to test the universality of free fall for neutrinos with data from the increased gamma-ray emission from two other blazars associated with TeV IceCube neutrino events, yielding improved constraints on the PPN parameter $\gamma$.

\section{\label{sec:method}Method of testing the WEP}

Let us begin with the PPN form of the Schwarzschild line element given by \citep{gravitation_book}
\begin{eqnarray}\label{PPN metric}
    ds^2 = - c^2 \left( 1-\frac{r_g}{r} \right) dt^2 + \left( 1+ \gamma \frac{r_g}{r} \right) \left[ dr^2 + r^2d\Omega^2 \right], 
\end{eqnarray}
where $ d\Omega^2 = d\theta^2+ \sin^2\theta d\phi^2 $ and c is the speed of light. In this framework, the PPN parameter $\gamma$ characterizes the amount of spacetime curvature generated by a unit rest mass. The quantity $r_g = \frac{2GM}{c^2}$ denotes the gravitational (Schwarzschild) radius associated with the mass $M$ of the central object, being the source of a spherically symmetric gravitational field. For $\gamma \neq 1$, the second term in the expression $\left( 1+ \gamma \frac{r_s}{r} \right)$  encodes potential deviations from WEP. For radial motion ($d\Omega = 0$) of particles propagating along null geodesics ($ds^2 = 0$) from the source (emitter) to the observer in the weak gravitational field of the central mass $M$, the resulting generalized expression for the Shapiro time delay, i.e. the delay experienced by a particle passing through a nonzero gravitational potential, extended to the more general case in which the PPN parameter $\gamma$ differs from unity, takes the following form:
\begin{equation}\label{Shapiro1}
    \delta t = \frac{GM}{c^3} (1+\gamma) \int_{r_e}^{r_o} \frac{dr}{r} 
\end{equation}
where  $r_e$ and $r_o$ denote the positions of the particle at the moments of emission and detection, respectively. If the distance between the source and the observer, as well as distance between the mass $M$ and the observer, are sufficiently large (as is typical for particles originating from distant astrophysical sources) the time delay given by Eq.~\ref{Shapiro1}  can be calculated in close analogy to the same problem for radar signals propagating near the Sun (see, e.g. \citealp{gravitation_book}). Adopting the source–mass M–detector geometry and notation used in \cite{Longo_1988} (see Fig.~1), the expression for the resulting time delay is now the following:
\begin{equation}\label{Shapiro2}
    \delta t = \frac{GM}{c^3} (1+\gamma) \ln \frac{ \left[(X_S + \sqrt{X_S^2 + b^2})(X_D + \sqrt{X_D^2 + b^2})\right]}{b^2}.
\end{equation}
Here $X_S$ is the distance between the source and mass $M$ projected on the line-of-sight, $X_D$ is the distance between the mass $M$ and detector projected on the line-of-sight and $b$ is the impact parameter.
From the perspective of an observer located at the detector, however, the Eq.~\ref{Shapiro2} is not directly suitable for constraining the PPN parameter $\gamma$ with observational data. The only measurable quantities are the distance to the mass $M$ ($r_D$), the distance to the source ($r_S$), and the angular separation $\theta$ between the direction to the source and the direction toward the mass $M$. Consequently, the expression for the Shapiro time delay in Eq.~\ref{Shapiro2} must be rewritten so that it depends solely on these observables.
Substituting $r_D = \sqrt{X_D^2 + b^2}$ and $r_S = \sqrt{X_S^2 + b^2} \sim X_S$ (the latter approximation is justified for cosmological sources such as blazars, for which $b << r_S$) into Eq.~\ref{Shapiro2} and expressing $\sin \theta$ and $\cos \theta$ in terms of the sides of the triangle defined by $b$, $X_D$, and $r_D$, a short algebraic manipulation yields the following operational formula for the generalized Shapiro time delay (see also \citealp{Wang_2016}):
\begin{equation}\label{Shapiro3}
    \delta t = \frac{GM}{c^3} (1+\gamma) \ln \left( \frac{2r_S}{r_D} \frac{1}{1-\cos \theta}\right).
\end{equation}
For $\gamma=1$, the above formula (Eq.~\ref{Shapiro3}) recovers the ordinary Shapiro time delay as predicted within general relativity.
Considering the motion of a particle (photon/neutrino) emitted from a blazar and detected by instruments on Earth, one must account for the gravitational influence of the cosmic structure surrounding the Earth at sufficiently large scales. In this context, the dominant contribution to the Shapiro delay arises from the gravitational potential of the Laniakea supercluster - the largest known structure that can be described using currently available distance and peculiar velocity data. Discovered in 2014 \citep{Tully_2014}, the Laniakea supercluster encompasses all major cosmic structures within the volume of $2 \times 10^6$ $(\text{Mpc}$ $h^{-1})^3$ 
\citep{DupuyCourtois_2023}, including numerous galaxy clusters (13 Abell clusters, including Virgo Cluster) and superclusters (Hydra Supercluster, Centaurus Supercluster), filaments (Pavo-Indus filament), and voids (Local Void, Sculptor Void) which reside within the local basin of attraction centered at the Right Ascension \text{R.A.} = $17^h30^m$ and the Declination \text{Dec.} = $-38^{\circ}46^{\prime}48^{\prime\prime}$ \citep{DupuyCourtois_2023} a direction coincident with the region known as the Great Attractor, widely interpreted as the Laniakea's gravitational center (see \citealp{DupuyCourtois_2023} and references therein). The Milky Way, as part of the Local Group, is located on the outskirts of the Laniakea supercluster. For our analysis, we adopt a rough mass estimate of $ 10^{17} M_{\odot} $ for Laniakea and an Earth-Laniakea center distance of $73.68$ Mpc $h^{-1}$ \citep{DupuyCourtois_2023}.

Classically, WEP implies that all particles respond identically to the spacetime curvature, regardless of their mass. Now let us assume that this is not the case. A convenient way to parameterize the deviations from WEP is to assign different values of the PPN parameter $\gamma$ to different types of particles. Under this assumption, a neutrino and a photon emitted simultaneously from the same source will not, in general, arrive at the detector at the same time, even though they propagate through the same gravitational field. 
In particular, from Eq.~\ref{Shapiro3}, the difference in arrival times between photons and neutrinos is:
\begin{equation}
    \Delta t \equiv \delta t_{\gamma} - \delta t_{\nu} = (\gamma_{\gamma} - \gamma_{\nu}) \delta t_S
\end{equation}
where $\delta t_S$ denotes the ordinary Shapiro time delay discussed earlier, i.e $\delta t_S = \frac{GM}{c^3} \ln \left( \frac{2r_S}{r_D} \frac{1}{1-\cos \theta}\right)$. Thus, comparing the arrival times of electromagnetic and neutrino signals provides a direct probe of potential WEP violations.

\section{\label{sec:test}Tests of the WEP with IceCube event and blazar}

As previously discussed, blazars - among the most powerful sources of electromagnetic radiation in the universe, situated at considerable distances - are particularly well suited for verifying fundamental physics, especially for WEP testing. To date, IceCube has detected multiple high-energy events spatially associated with various blazars, including TXS 0506+056 \citep{IceCube_2018a, Padovani_2018, Paiano_2018}, PKS 1424-418 \citep{Kadler_2016}, GB6 J1040+0617 \citep{Garrappa_2019}, 3HSP J095507.9+355101 \citep{Garrappa_2019}, PKS 1502+106 \citep{Rodrigues_2021}, and PKS 0735+178 \citep{Acharyya_2023}. Interactions within the jet are considered as a potential source of high-energy gamma rays and neutrinos observed from blazars. For TXS 0506+56 in particular, some models suggest that the high-energy gamma rays emitted alongside neutrinos interact with low-energy photons within the jet, leading to pair production and the development of electromagnetic cascades, with the energy ultimately being radiated by the source in the X-ray range \citep{Keivani_2018, Ansoldi_2018}. As the first high-energy neutrino event discovered and confirmed to be associated with a blazar, TXS 0506+56 represents the most promising candidate for testing the WEP. In addition to TXS 0506+056, we also include PKS 0735+178, which exhibits a smaller arrival‑time difference between neutrinos and photons, thereby enabling tighter constraints on the PPN parameter $\gamma$.\\

To ensure consistency in our calculations, we adopt a standard flat $\Lambda$CDM cosmological model with the same parameter values used in \cite{DupuyCourtois_2023}, i.e. $H_0 = 74.6$ kms$^{-1}$Mpc$^{-1}$ and $\Omega_m = 0.3$. Using these cosmological parameters, the corresponding proper comoving distance to the center of Laniakea is about 109 Mpc.

\subsection{\label{sec:src1}IC-170922A and TXS 0506+056}

On September 22, 2017, IceCube detected an upgoing muon induced by a neutrino (designated IceCube-170922A) with an energy of approximately 24 TeV, and a 56\% probability of being astrophysical in origin \citep{IceCube_2018b}. The parent neutrino's energy was estimated to be around 290 TeV. The neutrino's trajectory was found to align with the position of the TeV gamma-ray blazar TXS 0506+56. Following this detection, the Fermi-LAT and AGILE gamma-ray satellites reported an enhanced flare from the blazar within the energy range of 0.1 to 10 GeV \citep{Tanaka_2017, Acero_2015, Abdollahi_2017, Ajello_2017,Lucarelli_2019, IceCube_2018b}. About a week after the IceCube detection, the MAGIC atmospheric Cherenkov telescope also reported an excess of 100 GeV gamma rays, with a significance of $6\sigma$ relative to the expected background \citep{Mirzoyan_2017, Acciari_2022}. After evaluating three relevant neutrino/gamma-ray emission models, the association between IceCube-170922A and the gamma-ray flare was supported at better than $3\sigma$, suggesting that the correlation was unlikely to be coincidental \citep{IceCube_2018a, IceCube_2018b}. This event provided the first direct evidence that blazars can produce high-energy neutrinos and possibly ultra-high-energy cosmic rays.

TXS 0506+056 is a BL Lacertae-type blazar with a measured redshift of $z = 0.3365$ and its best-fit reconstructed position is at the Right Ascension R.A. = $5^h9^m43^s$ and the Declination Dec. = $+5^\circ43^{\prime}12^{\prime\prime}$ (J2000, 90\% containment) \citep{Padovani_2018,Paiano_2018}. The blazar exhibited an increase in gamma-ray emission in the GeV band beginning in April 2017, prior to the IceCube-170922A alert \citep{Abdollahi_2017, Tanaka_2017, IceCube_2018b}, suggesting that the maximum possible arrival time delay between the onset of the flare and the neutrino arrival is approximately 175 days. However, observations from Fermi-LAT and AGILE indicated that the peak of the high-energy gamma-ray flare occurred about 15 days before the neutrino event (see the left panel of Fig~3 in \cite{IceCube_2018b}, and references therein). Thus, assuming the neutrino was emitted at the time of the flare peak, the time delay between the arrival of the TeV neutrino and the gamma-ray photon would be approximately 15 days. Here, we take the 175-day delay as an upper limit on the time delay. 
Given the known redshift, the proper comoving distance to the blazar is estimated to be $r_S \simeq 1.246$ Gpc. The angle $\theta$ between the direction to the blazar and the direction to Laniakea's gravitational center is approximately equal to $33^{\circ}$. Putting $r_D = 109$ Mpc, one obtains the Shapiro time delay for relativistic particles emitted from the blazar under consideration and detected on Earth to be $\delta t_S \simeq 5.63$ days. This allows us to estimate the upper limits on possible deviations from WEP (i.e. on $\Delta \gamma = \gamma_{\gamma} - \gamma_{\nu}$) at the level of: $10^{-6}$ for $\Delta t = 175$ days, and $10^{-7}$ for $\Delta t = 15$ days and  $\Delta t = 7$ days. The exact numerical constraints derived in our analysis are summarized in the Table~\ref{tab:table1}.

\subsection{\label{sec:src2}IC-211208A and PKS 0735+178}

The intermediate-high-energy-peaked BL Lac object (IHBL) PKS 0735+178 is one of the brightest BL Lac objects in the sky \citep{Ciprini_2007, Hewitt_1987}. However, its spectrum is nearly featureless, making it difficult to determine its exact redshift. A lower limit of $z \geq 0.424$ was established by \cite{Carswell_1974} based on the observed possible absorption lines near $4000 \text{\AA}$, assuming this feature is identified as $\text{Mg-II}$ doublet. Given the challenges in determining redshift through the standard spectroscopic method due to the lack of emission features in the optical continuum of BL Lac objects, a complementary approach based on the so-called imaging redshift idea has been proposed \citep{Sbarufatti_2005}. This technique relies on deep imaging of the host galaxy, treating it as a standard candle. The redshift is then inferred using the distance modulus formula, under the assumption that the absolute magnitude distribution of BL Lac host galaxies follows an approximately Gaussian profile, with a mean value of $M^{host}_{R}=-22.8$ and a standard deviation of $\sigma = 0.5$ mag. Applying this method to PKS 0735+178 yields a preliminary redshift estimate of $z = 0.45 \pm 0.06$ \citep{Nilsson_2012}.
More recently, \cite{Falomo_2021} proposed a redshift of approximately $z \sim 0.65$, assuming the source is part of a group of faint galaxies detected near its position. It should be noted that even at $z = 0.424$, the corresponding radio and gamma-ray luminosities are among the highest known for objects of this type \citep{Prince_2023, Sahakyan_2022, Ciprini_2007}.

The position of PKS 0735+178 lies just above the $\sim13$ square degree 90\% localization error region of IceCube-211208A, a track-like event with an estimated energy of 172 TeV \citep{Acharyya_2023, Prince_2023, Sahakyan_2022}. It also falls within the larger error region (5.5 degrees, 50\% containment) of a cascade event detected by the Baikal neutrino telescope with an estimated energy of 43 TeV, which occurred 3.95 hours after the IceCube detection, with a $2.85\sigma$ chance coincidence probability \citep{Dzhilkibaev_2021}. Additionally, this source is reported to be within the error region of a GeV neutrino detected by the Baksan Underground Scintillation Telescope four days earlier, with a chance coincidence probability of $\sim3\sigma$ \citep{Petkov_2021}. A follow-up analysis by the KM3NeT underwater neutrino detector also revealed another neutrino, with an estimated energy of $\sim18$ TeV, detected on December 15, 2021, with a p-value of 0.14 associated with PKS 0735+178 \citep{Filippini_2022}.

Taking the lower redshift limit of $z = 0.424$, the proper comoving distance to PKS 0735+178 can be estimated to be $r_S \simeq 1.54$ Gpc. Using the blazar’s reported coordinates: the Right Ascension R.A. = $7^h38^m24^s$ and the Declination Dec. = $+17^\circ42^{\prime}36^{\prime\prime}$ (J2000) \citep{Goyal_2017}, one finds that the angle between the direction to the source and the direction to the Laniakea center is $\theta \simeq 25^{\circ}18^{\prime}$. This yields $\delta t_S \simeq 6.5\times10^{7}$ days for relativistic particles traveling from PKS 0735+178 to Earth. A similar value for the Shapiro time delay is obtained when using other redshift estimates for the source, which is why we do not consider those cases in our subsequent calculations. Assuming a time delay of $\Delta t = 4$ days between the arrival of the neutrino and the arrival of the photon (see the right panel of Fig.~1 in \citealp{Acharyya_2023}), the resulting upper limit on $\Delta \gamma$ for PKS 0735+178/IceCube-211208A is of the order of $10^{-8}$. This gives an improvement of roughly one order of magnitude over the constraints on potential WEP violation derived for the IC‑170922A/TXS 0506+056, primarily due to the shorter observed time delay between the neutrino and photon signals and the comparatively larger distance to the source.

\begin{table}[b]
\caption{\label{tab:table1}%
Upper limits on $\Delta \gamma \equiv \gamma_{\gamma} - \gamma_{\nu}$ with two sources under different conditions
}

\begin{ruledtabular}
\begin{tabular*}{0.2\textwidth}{lcc}
\textrm{Source}&
\textrm{Time delay [in days]}&
\textrm{$\Delta \gamma \equiv \gamma_{\gamma} - \gamma_{\nu}$}\\
\colrule
TXS 0506+056 & 7\ days & $ 1.24\times 10^{-7} $\\
  & 15\ days & $ 2.67\times 10^{-7} $\\
  & 175\ days & $ 3.11\times 10^{-6} $\\
PKS 0735+178 & 4 \ days & $6.13\times 10^{-8}$ \\
\end{tabular*}
\end{ruledtabular}
\end{table}

\section{\label{sec:discussion}Summary}

The universality of free fall - the Weak Equivalence Principle (WEP) - translates into an equal response of all particles to gravity as these particles follow the same trajectories in curved spacetime. In particular, within the parametrized post-Newtonian formalism (PPN), where the PPN $\gamma$ parameter characterizes the strength with which mass distorts the geometry of the spacetime. In general relativity, a theory essentially built upon WEP, one expects $\gamma = 1$ for all particles, regardless of their mass. This provides an opportunity to test WEP by constraining the PPN $\gamma$ parameter via the E\"{o}tv\"{o}s-type experiment for different particles. In this work, we investigated the possibility of deviations from WEP by searching for potential differences in the PPN $\gamma$ parameter between distinct particle types using gamma-ray blazars associated with high-energy neutrino events detected by IceCube. Compared to previous studies (e.g. \citealp{Wang_2016}), our findings indicate that the upper bounds on such possible differences (i.e. $\Delta \gamma \equiv \gamma_{\gamma} - \gamma_{\nu}$) for relativistic particles (photons and neutrinos) emitted from cosmological sources are significantly better, by one to three orders of magnitude. For comparison, we estimate the WEP constraint for the blazar PKS B1424-418 event studied by \cite{Wang_2016} under the same Laniakea gravitational potential adopted in this work. In their analysis, the gravitational potentials of the Virgo Cluster and the Great Attractor yielded limits of $\Delta \gamma \leq 3.4\times10^{-4}$ and $\Delta \gamma \leq 7\times10^{-6}$, respectively. Applying the same formalism as Eq.~\ref{Shapiro3}, we obtain a corresponding constraint of $\Delta \gamma \leq 3.2\times10^{-6}$ when the gravitational potential of the Laniakea supercluster is considered. This result shows that including the Laniakea potential can further tighten the astrophysical constraints on the Weak Equivalence Principle.

The constraints on the upper limit for the difference between the PPN parameters $\gamma$ derived from the TXS 0506+056 and PKS 0735+178 blazars are already quite precise, with $\Delta \gamma$ in the range of $10^{-8}$ to $10^{-6}$. The main obstacle to making these estimates more accurate is that we do not account for intrinsic time lags (the difference in the moments of emission between the neutrino and photon) originating at the source. Because we have no observational guidance regarding these intrinsic delays, the issue is inherently difficult to address. As a consequence, the constraints we obtain should be interpreted as conservative upper limits. Furthermore, since blazars lie at cosmological distances, uncertainties in distance estimates inferred from their redshifts can also influence the precision of our final results. We therefore expect that future measurements with improved distance determinations will yield more stringent constraints on potential WEP violation. Despite these limitations, our study provides a robust estimate, demonstrating that the universality of free fall for relativistic particles (and thus the WEP) is validated to a high degree of precision. With continued observations by IceCube and gamma‑ray detectors, additional neutrino–gamma‑ray blazar associations are expected to be identified. A larger sample of such events, combined with more accurate distance measurements and improved timing precision, will further tighten the constraints on the differences in the PPN parameter $\gamma$ across particle types, enabling an even more stringent test of the WEP. High energy astrophysics is entering its golden age. Capabilities of detecting very high-energy cosmic neutrinos with KM3NeT, in particular in ARCA detector, have already been demonstrated by the detection of $120\;PeV$ muon \citep{KM3Net_2025Nature}, which most probably originated from the interaction (with water) of a neutrino of even higher energy. Cherenkov Telescope Array Observatory (CTAO) is under  construction and its operational phase is approaching. Energies of very high-energy photons up to $300\; TeV$ \citep{CTA} will push the CTAO beyond the edge of the known electromagnetic spectrum and will open fascinating avenues to test fundamental physics including Lorentz Invariance Violations. With these large high-energy astrophysics facilities, the prospects of further constraining WEP violations are really promising.

\begin{acknowledgments}

This research has made use of data from the IceCube Neutrino Observatory, a facility funded by the U.S. National Science Foundation (NSF) and partner institutions. We thank the IceCube Collaboration for making the high-energy neutrino event data publicly available. 
We also acknowledge the Fermi-LAT and MAGIC collaborations for the gamma-ray observations of the blazars TXS~0506+056 and PKS~0735+178 used in this work. 

This work was supported by the National Natural Science Foundation of China (Projects 12541304, 12373040), the National Key R\&D Program of China (grant Nos. 2024YFA1611703 and 2024YFA1611700) and the Fundamental Research Funds for the Central Universities.

M.B. was supported by the Polish National Science Centre grant 2023/50/A/ST9/00579. M.B. and A.P-K gratefully acknowledges the support from COST Action CA21136 – “Addressing observational tensions in cosmology with systematics and fundamental physics (CosmoVerse)”.

\end{acknowledgments}

%
\facilities{IceCube, Fermi(LAT), MAGIC}

\software{astropy \citep{2013A&A...558A..33A, 2018AJ....156..123A, 2022ApJ...935..167A}}





\bibliography{sample701}{}
\bibliographystyle{aasjournalv7}



\end{document}